\documentclass[superscriptaddress,pra]{revtex4}

\usepackage[latin1]{inputenc}
\usepackage{graphicx}
\usepackage{amsmath}
\usepackage{amssymb}
\usepackage{dsfont}
\usepackage{graphicx}

\makeatletter
\def\erf{\mathop{\operator@font erf}\nolimits}
\makeatother

\newcommand\be{\begin{equation}}
\newcommand\ee{\end{equation}}

\begin{document}

\title{Useful transformations: from ion-laser interactions to master equations}
\author{R. Ju\'arez-Amaro$^{1}$, J.M. Vargas-Mart\'{\i}nez$^{2}$ and H. Moya-Cessa$^{2}$} \affiliation{${}^1$Centro de Investigaciones en
Optica, A.C.,
Loma del Bosque 115, Lomas del Campestre, Le\'on, Gto., Mexico,\\
\small ${}^2$Universidad Tecnol\'ogica de la Mixteca, Apdo.
Postal 71, 69000 Huajuapan de Le\'on, Oax., Mexico\\
\small $^{3}$INAOE, Coordinaci\'on de Optica, Apdo. Postal 51 y
216, 72000 Puebla, Pue., Mexico}

\begin{abstract}
We show  set of transformations which allow to obtain analytic
solutions in several quantum-optical problems. We start with the
ion-laser (time dependent) interaction, continue with the problem
of a slow atom interacting with a quantized field to end with a
master equation that describes losses. In all cases it is shown
that one may find useful transformations that simplify the
problems.
\end{abstract}
 \maketitle

\section{Introduction}
The purpose of this contribution is to show some transformations
that simplify Hamiltonians such that they may be treated in an
analytic form. We will study various problems, starting with the
ion-laser interaction, where we have shown that there exists a
time dependent  transformation that linearizes the Hamiltonian
with no approximations \cite{Moya-1}. We follow with the
interaction of a slow atom with a quantized field, the main result
of the paper. In this case the atom is affected by the mode shape
of the field \cite{Larson} and we show that the interaction may be
simplified as we pass from a three-body system to a two-body
effective interaction, again with no approximations. Finally we
analyze the case of the master equation (ME) that describes losses
for an anharmonic oscillator \cite{Milburn}, one can transform
such a ME to obtain a simpler equation where all the
superoperators commute, allowing its simple integration
\cite{Moya-2}.

\section{Ion-laser interaction}

We consider the Hamiltonian of a single ion (with unity mass
 trapped in a harmonic potential in interaction with
laser light in the (optical) rotating wave approximation
\cite{expl})

\begin{equation}
\hat{H} = \frac{1}{2} \left[ \hat{p}^2 + \nu^2(t) \hat{x}^2
\right] + \hbar\omega_{21}\hat{A}_{22} +\hbar\lambda(t)[
E^{(-)}(\hat{x},t)\hat{A}_{12}+ H.c.], \label{1}
\end{equation}
$\hat{A}_{ab}$ are the operators relating the different electronic
transitions (two-level) flip operator for the $|b\rangle
\rightarrow |a\rangle$ transition of frequency $\omega_{21}$,
respectively. $\nu(t)$ is the trap (time dependent) frequency,
$\lambda$ the electronic coupling matrix element, and
$E^{(-)}(\hat{x},t)$ the negative part of the classical electric
field of the driving field. The operators $\hat{x}$ and $\hat{p}$
are the position and momentum of the centre of mass of the ion. We
assume the ion driven by a laser field $ E^{(-)}(\hat{x},t)$
\begin{equation}
E^{(-)}(\hat{x},t)= E_{0}e^{-i(k\hat{x}-\omega t)}. \label{2}
\end{equation}
We want to solve the Schr\"odinger equation
\begin{equation}
i\hbar \frac{\partial |\xi(t)\rangle}{\partial t}=
\hat{H}|\xi(t)\rangle,
\end{equation}
in order to do this,  we make the transformation $|\phi\rangle =
\hat{T}(t) |\xi\rangle$, with \cite{fer}
\begin{equation}
\hat{T}(t)=e^{i\frac{\ln\{\rho(t)\sqrt{\nu_0}\}}{2\hbar}(\hat{x}\hat{p}+\hat{p}\hat{x})}
e^{-i\frac{\dot{\rho}(t)}{2\hbar\rho(t)}\hat{x}^2} \label{trans1}
\end{equation}
with $\rho(t)$ a function that obeys the Ermakov equation
\begin{equation}
\ddot{\rho}+\nu^{2}(t)\rho=\frac{1}{\rho^3}. \label{erma}
\end{equation}
 such that
we obtain the equation for $|\phi\rangle$
\begin{equation}
i\hbar\frac{\partial |\phi(t)\rangle}{\partial t}= \hat{\cal
H}|\phi(t)\rangle, \label{eq6}
\end{equation}
with the transformed Hamiltonian given by
\begin{equation}
\hat{\cal H} = \frac{1}{2\nu_0\rho^2} \left( \hat{p}^2 + \nu^2_0
\hat{x}^2 \right) + \hbar\omega_{21}\hat{A}_{22} +\hbar\Omega(t)
[e^{-i(k\hat{x}\rho(t)\sqrt{\nu_0}-\omega t)}\hat{A}_{12}+ H.c.],
\label{3}
\end{equation}
with $\Omega=\lambda E_0$. We consider that
$\omega_{21}=\omega+\delta$ where $\delta$ is the so-called
detuning. We transform to a frame rotating at $\omega$ by means of
the transformation $\hat{T}_{\omega}=e^{-i\omega t \hat{A}_{22}}$
to obtain the Hamiltonian $\hat{\cal H_{\omega}}
=\hat{T}_{\omega}\hat{\cal H} \hat{T}_{\omega}^{\dagger}$
($|\phi\rangle\rightarrow|\phi_{\omega}\rangle$)
\begin{equation}
\hat{\cal H_{\omega}}
=\hbar\tilde{\omega}(t)\left(\hat{n}+\frac{1}{2}\right)+\hbar\delta\hat{A}_{22}+
\hbar\Omega(t) [e^{-i(\hat{a}
+\hat{a}^{\dagger})\eta(t)}\hat{A}_{12}+ H.c.]. \label{4}
\end{equation}
where $\hat{n}=\hat{a}^{\dagger}\hat{a}$ with
\begin{equation}
\hat{a}=\sqrt{\frac{\nu_0}{2\hbar}}\hat{x}+i\frac{\hat{p}}{\sqrt{2\hbar\nu_0}},
\qquad
\hat{a}^{\dagger}=\sqrt{\frac{\nu_0}{2\hbar}}\hat{x}-i\frac{\hat{p}}{\sqrt{2\hbar\nu_0}}
\end{equation}
the annihilation and creation operators respectively.
$\tilde{\omega}(t)= 1/\rho^2$ is the characteristic frequency of
the time dependent harmonic oscillator. The time dependent
Lamb-Dicke parameter is written as
$\eta(t)=\eta_0\rho(t)\sqrt{\nu_0}$ with
$\eta_0=k\sqrt{\frac{\hbar}{2\nu_0}}$, where $k$ is the magnitude
of the wave vector of the driving field.

We will consider now the resonant interaction ($\delta=0$).
Passing to a frame rotating at the frequency $\tilde{\omega}(t)$
we may get rid off the harmonic oscillator term in (\ref{4}) to
end up with the (time dependent) interaction Hamiltonian
\begin{equation}
\hat{\cal H}(t) =\hbar\Omega(t)
[e^{-i(\hat{a}e^{-i\int\tilde{\omega}(t')dt'}
+\hat{a}^{\dagger}e^{i\int\tilde{\omega}(t')dt'})\eta(t)}\hat{A}_{12}+
H.c.]. \label{timedep}
\end{equation}

\subsection{Linearizing the system} Finally we make the
transformation $|\psi\rangle=\hat{R}(t)|\phi\rangle$ with (see
\cite{Moya} for the time independent case)
\begin{equation}
\hat{R}(t)=e^{\frac{\pi}{4}(\hat{A}_{21}-\hat{A}_{12})}e^{-i\frac{\eta(t)}{2}(\hat{a}+\hat{a}^{\dagger})(\hat{A}_{22}-\hat{A}_{11})}
\label{trans2}
\end{equation}
to obtain
\begin{equation}
i\hbar\frac{\partial |\psi\rangle}{\partial t}
=\hbar\left\{\tilde{\omega}(t)\hat{n}+ \Omega
(\hat{A}_{22}-\hat{A}_{11}) +
\left(\frac{\delta}{2}+i[\hat{a}\beta(t)-\hat{a}^{\dagger}\beta^*(t)]\right)(\hat{A}_{12}+\hat{A}_{21)}
\right\}|\psi\rangle, \label{final}
\end{equation}
with $\beta(t)=\frac{\eta(t)\tilde{\omega}}{2}-i\dot{\eta}(t)/2$
and we have disregarded the term $\tilde{\omega}/2$ as it would
add an overall phase. A method to solve JC-like interactions with
time dependent parameters has been published by Shen {\it et al.}
\cite{shen}.
\subsection{Many ions} We  generalize the
transformation that allows to linearize the ion-laser Hamiltonian
\cite{Moya} in an exact form for two different interactions,
namely, for the case of many ions in interaction with laser fields
\cite{Molmer}, and for an ion vibrating in two dimensions
interacting with a laser field. This linearization has been shown
to be important for instance in the implementation of fast gates
in ion-laser interactions \cite{Plenio}. This is possible, as more
regimes may be studied with such a linearization: high intensity,
low intensity and middle intensity regimes.

In particular we will show that in the case of many ions, the
transformation produces a term which correspond to a dipole-dipole
interaction.

Ions in a linear trap interacting with a laser field may be
described by the Hamiltonian \cite{Molmer}
\begin{equation}
{H}_M = \nu a^{\dagger}a + \frac{\delta}{2}\sum_{j}\sigma_{zj} +
\sum_{j} \Omega_j(\sigma_{+j}e^{i\eta_j(a^{\dagger}+a)}+ H.c.)
\label{many}
\end{equation}
where $\nu$ is the frequency of the vibration, $a^{\dagger}$ and
$a$ are the creation and annihilation operators of the quantized
oscillator, $\delta$ is the detuning between the transition
frequency of the internal states of the ion, $\omega_{eg}$, and
the laser frequency $\omega_L$, and $\Omega_i$ is the resonant
Rabi frequency of the $i$'th ion in the laser field. The
exponentials account for the position dependence laser-field and
the recoil of the ions upon absorption of a photon. The positions
of the ions $x_i$ are replaced by ladder operators
$kx_i=\eta_i(a^{\dagger}+a)$, where the Lamb-Dicke parameter
$\eta_i$ represents the ration between the ionic excursions within
the vibrational ground state wavefunction and the wavelenght of
the exciting radiation. We can linearize the Hamiltonian
(\ref{many}) via the transformation \cite{Moya}
\begin{equation}
T_{M}=\prod_{j}e^{\frac{\pi}{4}(\sigma_{+j}-\sigma_{-j})}e^{-i\eta_j\sigma_{zj}(a^{\dagger}+a)/2}
\end{equation}
which gives the Hamiltonian
\begin{eqnarray}
\nonumber
\mathcal{H}_M &=& THT^{\dagger} \\
&=& \nu a^{\dagger}a - \frac{\delta}{2}\sum_{j}\sigma_{xj}
+\sum_{j}\Omega_j\sigma_{zj}+i\sum_{j}\frac{\eta_j\nu(a-a^{\dagger})}{2}\sigma_{xj}
+ \sum_{j,k}\frac{\eta_j\eta_k}{4}\sigma_{xj}\sigma_{xk}
\end{eqnarray}
Here $\sigma_{xj}=\sigma_{+j}+\sigma_{-j}$. Note that the
transformation, besides linearizing the Hamiltonian, produces an
ion-ion (dipole) interaction.

\subsection{Two-dimensional vibration} An ion vibrating in
two-dimensions has the Hamiltonian
\begin{equation}
H_{2d}=\nu_x a_x^{\dagger}a_x+\nu_y a_y^{\dagger}a_y +
\frac{\delta}{2}\sigma_z +\Omega(\sigma_+e^{i\eta_x(
a_x^{\dagger}+a_x)}e^{i\eta_y( a_y^{\dagger}+a_y)}+ H.c.)
\label{2d}
\end{equation}
with the transformation
\begin{equation}
T_{2d}=e^{\frac{\pi}{4}(\sigma_{+}-\sigma_{-})}e^{-i[\eta_x(a_x^{\dagger}+a_x)
+\eta_y(a_y^{\dagger}+a_y)]\sigma_{z}/2}
\end{equation}
we can cast the Hamiltonian (\ref{2d}) into the linearized
Hamiltonian
\begin{equation}
\mathcal{H}_{2d}=\nu_x a_x^{\dagger}a_x+\nu_y a_y^{\dagger}a_y -
\frac{\delta}{2}\sigma_x
+\Omega\sigma_x+\frac{i\nu}{2}\sigma_x[\eta_x(a_x-a_x^{\dagger})+\eta_y(a_y-a_y^{\dagger})]
\end{equation}
where we have disregarded a constant term. The Hamiltonian above
looks like a two-mode quantized-field-atom interaction.
\section{Slow atom interacting with a quantized field}
Here we show how a three body problem may be reduced to a two body
problem via a transformation, we treat the problem of a slow atom
interacting with a quantized field. Because the slowness of the
atom, the field mode-shape affects the interaction. We can write
down the Hamiltonian describing a single two-level atom passing an
electromagnetic field confined to a cavity. Inn addition to the
Jaynes-Cummings Hamiltonian, we have to add the energy of the free
atom and the spatial variation it feels from the cavity, the
Hamiltonian reads
\begin{equation}
H=\frac{p^2}{2}+\omega\hat{n}+\frac{\omega_0}{2}\sigma_{z}+g(x)(\hat{a}\sigma_{+}
+\hat{a}^{\dagger}\sigma_{-}), \label{slow}
\end{equation}
on resonance, we can pass to the interaction picture Hamiltonian
\begin{equation}
H_I=\frac{p^2}{2}+g(x)(\hat{a}\sigma_{+}+\hat{a}^{\dagger}\sigma_{-}),
\end{equation}
we use the $2\times 2$ notation for the Pauli spin matrices and
write the interaction Hamiltonian as (see \cite{Moya-2})
\begin{equation}
\hat{H}_I= \frac{p^2}{2}+g(x)\hat{T}^{\dagger} \left(
\begin{array}{cc}
0 & \sqrt{\hat{n}+1}
\\  \sqrt{\hat{n}+1}& 0
\end{array}
\right) \hat{T} \label{INT}
\end{equation}
where a non unitary transformation $\hat{T}$ has been used. We
define $\hat{T}$ as
\begin{equation}\hat{T}=\left(
\begin{array}{cc}
1 & 0
\\ 0 & \hat{V}
\end{array}
\right)
\end{equation}
Note that $\hat{T}\hat{T}^{\dagger}=1$ but
$\hat{T}^{\dagger}\hat{T}=1-\rho_{g,v}$ with
\begin{equation}\rho_{g,v}=\left(
\begin{array}{cc}
0 & 0
\\ 0 & |0\rangle\langle 0|
\end{array}
\right)
\end{equation}
We can use the definitions above to rewrite (\ref{slow}) as
\begin{equation}
\hat{H}_I=(\hat{T}^{\dagger}\hat{T}+\rho_{g,v})
\frac{p^2}{2}(\hat{T}^{\dagger}\hat{T}+\rho_{g,v})+g(x)\hat{T}^{\dagger}
\left(
\begin{array}{cc}
0 & \sqrt{\hat{n}+1}
\\  \sqrt{\hat{n}+1}& 0
\end{array}
\right) \hat{T} \label{INT}
\end{equation}
By noting that $\hat{T}\rho_{g,v}=0$ we rewrite the above equation
as
\begin{equation}
\hat{H}_I=\hat{T}^{\dagger} \frac{p^2}{2}\hat{T}
+\frac{p^2}{2}\rho_{g,v}+g(x)\hat{T}^{\dagger} \left(
\begin{array}{cc}
0 & \sqrt{\hat{n}+1}
\\  \sqrt{\hat{n}+1}& 0
\end{array}
\right) \hat{T}
\end{equation}
where we have used that $\rho_{g,v}^2=\rho_{g,v}$. Finally we
factorize the transformation operators in the Hamiltonian above to
obtain
\begin{equation}
\hat{H}_I=\hat{T}^{\dagger}( \frac{p^2}{2}
+g(x)\sigma_{x}\sqrt{\hat{n}+1})\hat{T} +\frac{p^2}{2}\rho_{g,v}
\end{equation}
Note that $[\hat{T}^{\dagger}( \frac{p^2}{2}
+g(x)\sigma_{x}\sqrt{\hat{n}+1})\hat{T},\frac{p^2}{2}\rho_{g,v}]=0$
so that the evolution operator for the Hamiltonian above is given
by
\begin{equation}
\hat{U}_I(t)=e^{-i\hat{T}^{\dagger}( \frac{p^2}{2}
+g(x)\sigma_{x}\sqrt{\hat{n}+1})\hat{T}t}e^{-i
\frac{p^2}{2}\rho_{g,v}t}
\end{equation}
to obtain the first exponential we can do Taylor series, and we
note that the powers of the argument are simply
\begin{equation}
[\hat{T}^{\dagger}( \frac{p^2}{2}
+g(x)\sigma_{x}\sqrt{\hat{n}+1})\hat{T}]^k=\hat{T}^{\dagger}(
\frac{p^2}{2} +g(x)\sigma_{x}\sqrt{\hat{n}+1})^k\hat{T}  , \qquad
k\ge 1
\end{equation}
such that
\begin{equation}
e^{-i\hat{T}^{\dagger}( \frac{p^2}{2}
+g(x)\sigma_{x}\sqrt{\hat{n}+1})\hat{T}t}= \hat{T}^{\dagger}e^{-i(
\frac{p^2}{2} +g(x)\sigma_{x}\sqrt{\hat{n}+1})t}\hat{T}
+\rho_{g,v}. \label{evol}
\end{equation}
Note that  the evolution operator in (\ref{evol}) is  effectively
the interaction of two systems, as it is written in a form in
which the filed operators commute, unlike the Hamiltonian
(\ref{slow}), where all the operator involved do not commute.
\section{Master Equations}
Now we turn our attention to the superoperator solution of master
equations for more quantum optical systems, namely a dissipative
cavity filed with a Kerr medium \cite{Milburn}, master equation
describing phase sensitive processes \cite{Scully} and parametric
down conversion \cite{Walls}. Usually these equations are solved
by transforming them to Fokker-Planck equations \cite{Risken}
which are partial differential equations for quasiprobability
distribution functions typically the Glauber-Sudarshan
$P$-function and the Husimi $Q$-function. Another usual approach
is  to solve system-environment problems is through the use of
Langevin equations, this is stochastic differential equations that
are equivalent to the Fokker-Planck equation \cite{Gardiner}.

These approaches to the problem makes it usually difficult to
apply the solutions to an arbitrary initial field in contrast with
the superoperator techniques where it is direct the application to
an initial wave function. We have used this feature in the former
Section where we have exploited this fact to obtain reconstruction
mechanisms that allowed us to obtain information on the state of
the quantized electromagnetic field via quasiprobability
distribution functions.

\subsection{Kerr medium}
Before applying superoperator methods in the solution of the above
equation, let us show how it may be casted into a Fokker-Planck
equation. In order to do this one writes the density matrix in
terms of the Glauber-Sudarshan $P$-function, $\hat{\rho}=1/\pi\int
P(\alpha)|\alpha\rangle\langle\alpha|d^2\alpha$. Noting that the
creation and annihilation operators have the following relations
with the coherent state density matrix \cite{Louissel}
\begin{equation}
\hat{a}^{\dagger} |\alpha\rangle\langle \alpha|
=\left(\frac{\partial}{\partial \alpha} +
\alpha^*\right)|\alpha\rangle\langle \alpha|, \label{alfaparcial}
\end{equation}
\begin{equation}
 |\alpha\rangle\langle \alpha|\hat{a}
=\left(\frac{\partial}{\partial \alpha^*} + \alpha
\right)|\alpha\rangle\langle \alpha|, \label{alfaparcialc}
\end{equation}
we can obtain the following correspondence
\begin{equation}
\hat{a} \hat{\rho}\rightarrow \alpha P(\alpha), \qquad
\hat{a}^{\dagger} \hat{\rho} \rightarrow \left(\alpha^* -
\frac{\partial}{\partial \alpha} \right)P(\alpha),
\end{equation}
and
\begin{equation}
 \hat{\rho}\hat{a}^{\dagger}\rightarrow \alpha^* P(\alpha),
\qquad
 \hat{\rho}\hat{a} \rightarrow \left(\alpha -
\frac{\partial}{\partial \alpha^*} \right)P(\alpha).
\end{equation}
 In this form, whenever a creation or annihilation operator
occurs in the master equation, we can translate this into a
corresponding operation on the Glauber-Sudarshan $P$-function. The
equation that results is a Fokker-Planck equation \cite{Risken}
\begin{equation}
\frac{\partial P(\alpha,t)}{\partial t}=\left[\gamma \left(
\frac{\partial }{\partial \alpha} \alpha + \frac{\partial
}{\partial \alpha^*} \alpha^* \right)+2\gamma \bar{n}
\frac{\partial^2 }{\partial \alpha \partial \alpha^*}
\right]P(\alpha,t).
\end{equation}
This equation is equivalent to the stochastic differential
equation \cite{Gardiner}
\begin{equation}
\frac{d \alpha}{d t}=\gamma  \alpha + \sqrt{2\gamma \bar{n}}
\xi(t), \label{Lange}
\end{equation}
and the corresponding complex conjugate equation. The quantity
$\xi(t)$ is a white noise fluctuating force with the following
correlation properties
\begin{equation}
\langle \xi(t) \rangle=0,\qquad \langle \xi(t)
\xi^*(t)\rangle=\delta(t-t')\qquad \langle \xi(t)
\xi(t)\rangle=\langle \xi^*(t) \xi^*(t)\rangle=0
\end{equation}
Equation (\ref{Lange}) is usually called a Langevin equation, and
also as the "Stratonovich form" of the Fokker-Plank equation (see
for instance \cite{Strat,Dutralibro}). The Fokker-Planck equation
can be also obtained from the Kramers-Moyal \cite{Kramers,Moyal}
expansion, a Langevin equation that does not stop after second
order derivatives (see for instance \cite{Puri}). The master
equation for a Kerr medium in the Markov approximation and
interaction picture has the form \cite{Milburn}
\begin{equation}
\frac{d\hat{\rho}}{dt}=-i\chi[\hat{n}^2,\hat{\rho}]+2\gamma\hat{a}\hat{\rho}\hat{a}^{\dagger}
-\gamma \hat{a}^{\dagger}\hat{a}\hat{\rho} -\gamma
\hat{\rho}\hat{a}^{\dagger}\hat{a}. \label{kerr}
\end{equation}
Milburn and Holmes  \cite{Milburn} solved this equation  by
changing it to a partial differential equation for the
$Q$-function and for an initial coherent state. We can have a
different approach to the solution by again using superoperators.
If we define
\begin{equation}
\hat{Y}\hat{\rho}=-i\chi[\hat{n}^2,\hat{\rho}]
\end{equation}
we rewrite (\ref{kerr}) as
\begin{equation}
\frac{d\hat{\rho}}{dt}=(\hat{Y}+\hat{J}+\hat{L})\hat{\rho},
\end{equation}
where the superoperators $\hat{J}$ and $\hat{L}$ are defined as
\begin{equation}
\hat{J}\hat{\rho}=2\gamma\hat{a}\hat{\rho}\hat{a}^{\dagger},
\qquad \hat{L}\hat{\rho}=-\gamma
\hat{a}^{\dagger}\hat{a}\hat{\rho} -\gamma
\hat{\rho}\hat{a}^{\dagger}\hat{a}.
\end{equation}

 Now we use the transformation
\begin{equation}
\hat{\tilde{\rho}}=\exp[(\hat{Y}+\hat{L})t]\hat{\rho}
\end{equation}
to obtain
\begin{equation}
\frac{d\hat{\tilde{\rho}}}{dt}=\exp[-i\chi \hat{R}t- 2\gamma
t]\hat{J}\hat{\tilde{\rho}}, \label{Eq-milb}
\end{equation}
with
\begin{equation}
\hat{R}\hat{\tilde{\rho}}=2(\hat{n}\hat{\tilde{\rho}} -
\hat{\tilde{\rho}}\hat{n}).
\end{equation}
In arriving to  equation  (\ref{Eq-milb}) we have used the formula
$e^{y\hat{A}}\hat{B}e^{-y\hat{A}}=\hat{B}+y[\hat{A},\hat{B}]+y^2/2![\hat{A},[\hat{A},\hat{B}]]
+\dots $ for $y$ a parameter and $\hat{A}$  and $\hat{B}$
operators. We have also used the commutation relation
\begin{equation}
[\hat{Y},\hat{J}]\hat{\rho} =2i\chi\hat{R}\hat{J}\hat{\rho}.
\end{equation}

Now it is easy to show that $\hat{R}$ and $\hat{J}$ commute, so
that we can finally find the solution to equation (\ref{kerr}) as
\begin{equation}
\hat{\rho}(t)=e^{\hat{Y}t}e^{\hat{L}t}\exp[e^{\frac{-i\chi
\hat{R}t- 2\gamma t -1}{-i\chi \hat{R}- 2\gamma
}}\hat{J}]\hat{\rho}(0).
\end{equation}
Note that the above solution may be applied easily to any initial
density matrix:
\begin{eqnarray}
\nonumber \hat{\rho}(t)&=&\sum_{k,n,m=0}^{\infty}
\hat{\rho}_{n+k,m+k}(0)e^{-i\chi t(n^2-m^2)-\gamma
t(n+m)}\sqrt{\frac{(n+k)!(m+k)!}{n!m!}} \\
&&\left[\frac{1-e^{-2i\chi t(n-m)-2\gamma t}}{2i\chi (n-m)+2\gamma
}\right]^k \frac{(2\gamma)^k}{k!}|n\rangle \langle n|,
\end{eqnarray}
where $\hat{\rho}_{n,m}(0)$ are the (Fock) matrix elements of the
initial density matrix.
\section{Conclusions}
We have shown the importance of looking for transformations that
simplify Hamiltonians before trying to solve a given system. In
particular, we have shown how a $3$-body system can be reduced to
a $2$-body system, in the case of a slow atom interacting with a
quantized field. We have given the most complete solutions for the
ion-laser interaction in several cases: time dependent case,
several ions, ions vibrating in two dimensions.


\begin{thebibliography} {99}
\bibitem{Moya-1} J.M. Vargas-Mart\'{\i}nez and H. Moya-Cessa, {\it J. of Optics} B {\bf 6}, S618
(2004).
\bibitem{Larson} J. Larson and S. Stenholm, {\it Phys. Rev.} A {\bf 73}, 033805
(2006).
\bibitem{Milburn}G. J. Milburn and C. A. Holmes,  {\it Phys. Rev. Lett.} {\bf 56}, 2237 (1986)
\bibitem{Moya-2}  H. Moya-Cessa, {\it Phys. Rep.} {\bf 432}, 1 (2006).
\bibitem{expl} R.L. de Matos Filho and W. Vogel, {\it Phys. Rev. Lett.} {\bf 76}, 608
(1996);  R.L. de Matos Filho and W. Vogel, {\it Phys. Rev.} A {\bf
54}, 4560 (1996); H. Moya-Cessa, S. Wallentowitz, and W. Vogel,
{\it Phys. Rev.} A {\bf 59}, 2920 (1999); H. Moya-Cessa and P.
Tombesi, {\it Phys. Rev.} A {\bf 61}, 025401(2000).
\bibitem{fer} M. Fern\'andez Guasti and H. Moya-Cessa, {\it J. of Phys.} A {\bf  36}, 2069
(2003); H. Moya-Cessa and M. Fern\'andez Guasti, {\it Phys. Lett.}
A {\bf 311}, 1 (2003).
\bibitem{Moya} H. Moya-Cessa, A.
Vidiella-Barranco, J.A. Roversi, S.D. Freitas and S.M. Dutra,
Phys. Rev. A {\bf 59}, 2518 (1999); H. Moya-Cessa, D. Jonathan and
P.L. Knight, J. of Mod. Optics {\bf 50}, 265 (2003).
\bibitem{shen} J. Q. Shen, H.Y. Zhu P. and Chen,  {\it Eur. Phys. J.} D {\bf 23}
305 (2003).
\bibitem{Molmer} A. Sorensen and K. Molmer,  Phys. Rev. A {\bf 62}, 022311 (2000).
\bibitem{Plenio} D. Jonathan, M.B. Plenio and P.L. Knight, Phys. Rev. A {\bf 62}, 042307 (2000).
\bibitem{Scully} M.O. Scully, M.S. Zubairy, {\it Quantum Optics,} (Cambridge University Press, New York, NY,
1997).
\bibitem{Walls} D.F.Walls, G.J. Milburn,{\it Quantum Optics} (Springer, New York, 1994).
\bibitem{Risken} H. Risken, {\it The Fokker-Planck Equation: Methods of Solutions and Applications}, (Springer, Berlin, 1984).
\bibitem{Gardiner} C.W. Gardiner,  {\it Handbook of Stochastic Methods}, 2nd ed., (Springer, Heidelberg,
1990).
\bibitem{Louissel} W.H. Louisell,  {\it Quantum Statistical Properties of Radiation,} (Wiley, NewYork, 1973).
\bibitem{Strat} R.L. Stratonovich, {\it Introduction to the Theory of Quantum Noise,} (Gordon and Breach, NewYork, London, 1963).
\bibitem{Dutralibro} S.M. Dutra, {\it Cavity Quantum Electrodynamics}, (Wiley-Interscience, New York, 2005).
\bibitem{Kramers} H.A. Kramers, {\it Physica} {\bf 7}  284 (1940).
\bibitem{Moyal} J.E. Moyal, {\it J.R. Stat. Soc.} {\bf 11}  151 (1949).
\bibitem{Puri} R.R. Puri, {\it Mathematical Methods of Quantum Optics,} (Springer, Berlin, 2001).
\end{thebibliography}
\end{document}